\documentclass[pra,aps,twocolumn,showpacs,floatfix]{revtex4}
\usepackage{amsmath}

\usepackage{simon}

\usepackage{graphicx}

\begin{document}

\title{Vacuum Squeezing in Atomic Media via Self-Rotation}

\author{A. B. Matsko}
\author{I. Novikova}
\author{G. R. Welch}
\affiliation{
    Department of Physics and Institute for Quantum Studies,
    Texas A\&M University,
    College Station, Texas 77843-4242.%
}

\author{D. Budker}
\author{D. F. Kimball}
\author{S. M. Rochester}
\affiliation{
    Department of Physics, University of California, Berkeley,
    California 94720-7300
}

\date{\today}

\begin{abstract}
When linearly polarized light propagates through a medium in which elliptically polarized
light would undergo self-rotation, squeezed vacuum can appear in the orthogonal
polarization. A simple relationship between self-rotation and the degree of vacuum
squeezing is developed. Taking into account absorption, we find the optimum conditions
for squeezing in any medium that can produce self-rotation. We then find analytic
expressions for the amount of vacuum squeezing produced by an atomic vapor when light is
near-resonant with a transition between various low-angular-momentum states. Finally, we
consider a gas of multi-level Rb atoms, and analyze squeezing for light tuned near the
$D$-lines under realistic conditions.
\end{abstract}

\pacs{42.50.Dv, 42.50.Gy, 42.50.Ar, 32.60.+i}

\maketitle

\section{Introduction}

Quantum fluctuations are the main factor limiting the precision of many measurements and
the signal-to-noise ratio in optical communication. However, shot-noise is not a
fundamental restriction on the precision of an optical measurement if nonclassical states
of light are used~\cite{Caves81,Walls83,Loudon87,Leuchs88,Walls95,Scully97}. For example,
squeezed vacuum has been used to improve the performance of optical interferometers
beyond the shot-noise limit~\cite{Xaio,Grangier}.  A number of techniques for producing
squeezed states of light using nonlinear optics have been developed~(see, e.g.,
Ref.~\cite{JOSASqueeze} and references therein).

It has been known for some time that if linearly polarized light propagates through a
medium that can produce nonlinear self-rotation (SR) of elliptically polarized light
(see, e.g., Refs.~\cite{MakerTerhune64,budker01,boyd92,novikova00ol,novikova'01} and
references therein), the light will be in a squeezed state after traversal of the medium
if losses in the medium are sufficiently small. A detailed theoretical study of this
effect for a generic Kerr medium without absorption was carried out in
Ref.~\cite{Tanas83}. It can be difficult to achieve efficient squeezing in a realistic
Kerr medium because of the absorption and scattering of light. Nonetheless, squeezed
electromagnetic vacuum was recently produced in a nonbirefringent optical fiber in which
elliptically polarized light undergoes SR~\cite{Haus98}.

In this paper, we derive a simple formula for the amount of squeezing generated by a
medium which exhibits self-rotation, including the effects of absorption (which acts to
diminish the squeezing). In atomic vapors, one can take advantage of resonant enhancement
of self-rotation (as well as coherence and interference
effects~\cite{harris97phys.today,hemmer95ol,lukin98prl,zibrov99prl,lukin99prl,fleischhauer00prl,schmidt96ol,gheri99pra,harris98prl})
to create a medium in which there is large SR and small absorption. Employing
density-matrix calculations, we analyze vacuum squeezing of resonant light in some
low-angular-momentum atomic systems and in atomic rubidium. We find that under readily
achievable experimental conditions, vacuum squeezing of about 8~dB can be expected.

\section{Relationship between self-rotation and vacuum squeezing}

Before embarking on the calculations, we present a qualitative explanation of why
self-rotation leads to squeezing. Suppose that we have a strong linearly polarized (along
$\mb{\hat{y}}$) light field propagating in the $\mb{\hat{z}}$ direction through a medium
in which elliptically polarized light would undergo SR. Also suppose there is a weak
co-propagating $x$-polarized light field. In general, the resultant field is elliptically
polarized and the principal axis of the polarization ellipse will rotate upon propagation
through the medium.  This rotation projects a portion of the strong field along
$\mb{\hat{x}}$. Depending on the relative phase between the orthogonally-polarized strong
and weak input fields, the $x$-polarized output field can be amplified or attenuated
compared to the weak input field (phase-sensitive gain or loss). In the latter case, we
have a negative feedback mechanism which reduces the field along $\mb{\hat{x}}$. This
analysis applies even when the weak input field is solely due to vacuum fluctuations.

To describe squeezing that accompanies propagation of light through a nonlinear
SR-medium, we first introduce a phenomenological description of SR. The interaction of
light of ellipticity $\ge$ with an atomic medium of length $\ell$ can induce circular
birefringence and linear dichroism in the medium. This results in rotation by an angle
$\gv$ of the principal axis of the light polarization ellipse. Mechanisms for SR have
been discussed in, e.g., Ref.\ \cite{budker01}. For optically thin media and small
initial ellipticity $\ge(0)\ll1$, the principal axis of the polarization ellipse rotates
by an angle
\begin{equation}
    \gv=g\ge(0)\ell, \label{g}
\end{equation}
where we have defined an SR parameter $g$ that, for a given atomic medium, depends only
on the incident light intensity and frequency. Because the initial ellipticity of the
light field is small, $\gv\ll1$. We also assume that $\ge$ does not change significantly
as light propagates through the medium (i.e., $\epsilon$ is independent of $z$, so that
$\epsilon(0)=\epsilon(\ell)=\epsilon$), an assumption that we will later show is
justified in concrete examples considered in Secs.~\ref{SimpleSystems} and
\ref{RbDlines}.

The absorption of light by the medium is characterized by the parameter $\ga$; the light
intensity $I(z)$ as a function of the distance $z$ the light has propagated through the
medium is given by $I(z)=I(0)\exp\prn{-\ga{z}}$.

The parameters $g$ and $\ga$ can be measured or calculated (see
Refs.~\cite{budker01,boyd92,novikova00ol,novikova'01} and Secs.~\ref{DMcalc} and
\ref{SimpleSystems}); knowledge of these two parameters allows one to calculate the
amount of vacuum squeezing produced by a given system, in the following manner. Consider
a monochromatic light field $\mb{E}(z,t)$ propagating in the $\uv{z}$ direction
represented by
\begin{equation}
\begin{split}
    \mb{E}(z,t) & = E_x(z,t) \uv{x} + E_y(z,t) \uv{y}\\
    E_x(z,t) & = \mc{E}_x(z)\cos\sbr{kz-\go{t}+\gf(z)}\\
    E_y(z,t) & = \mc{E}_y(z)\cos\prn{kz-\go{t}}
\end{split}
\end{equation}
where $\mc{E}_x(z)$ and $\mc{E}_y(z)$ are the real positive amplitudes of the $x-$ and
$y-$polarized components of the electromagnetic field, $\go$ is the light frequency,
$k=\go/c$ is the vacuum wave number, and $\gf(z)$ is the relative phase between the two
components. The light field can also be written in terms of its positive- and
negative-frequency components, $E_{x,y}^+(z,t)$ and $E_{x,y}^-(z,t)$, where
\begin{equation}
\begin{split}
    \mb{E}^+(z,t)
        &=\frac{\mc{E}_x(z)}{2} e^{i\sbr{kz-\go{t}+\gf(z)}} \uv{x}
         +\frac{\mc{E}_y(z)}{2} e^{i\prn{kz-\go{t}}}        \uv{y}\\
    \mb{E}^-(z,t)
        &=\frac{\mc{E}_x(z)}{2} e^{-i\sbr{kz-\go{t}+\gf(z)}}\uv{x}
         +\frac{\mc{E}_y(z)}{2} e^{-i\prn{kz-\go{t}}}      \uv{y}~.
\end{split}
\end{equation}

The ellipticity $\ge$ of the light field is given by (see, e.g., Ref.\ \cite{Huard97})
\begin{equation}\label{ExactEpsilon}
    \ge =
        \frac{1}{2}
        \arcsin \frac{i\prn{E_x^-E_y^+ - E_y^-E_x^+}}{\abs{E_x}^2 + \abs{E_y}^2}.
\end{equation}
Assuming that $\mc{E}_y(z)\gg\mc{E}_x(z)$, we have for the ellipticity $\ge(z)$,
\begin{equation}\label{ClassEllip}
    \ge(z) \approx \frac{\mc{E}_x(z)}{\mc{E}_y(z)} \sin\gf(z).
\end{equation}

First we consider an SR-medium of length $\ell$ where absorption is negligible ($\ga\ell
\approx 0$), and relate the optical fields at the output to the input fields, using
Eq.~(\ref{g}).  For simplicity, we deal only with the positive-frequency components of
the optical electric field. The output fields $E^+_x(\ell)$ and $E^+_y(\ell)$ are then
given by
\begin{equation}
   \sbr{
     \begin{array}{c}
     E^+_x(\ell) \\
     E^+_y(\ell) \\
     \end{array}
   }
   \approx
   \sbr{
     \begin{array}{cc}
     1 & \gv \\
     -\gv & 1 \\
     \end{array}
   }
   \sbr{
     \begin{array}{c}
     \mc{E}_x(0)e^{i\gf(0)} \\
     \mc{E}_y(0) \\
     \end{array}
   }
   e^{i\prn{k\ell-\go{t}}}.
\end{equation}
The positive-frequency part of the $x$-polarized component of the output field is given
by
\begin{equation}\label{ClassEx}
    E^+_x(\ell) \approx
        \sbr{
            \mc{E}_x(0) e^{i\gf(0)}
            + g\ell\ge(0) \mc{E}_y(0)
        }
        e^{i(k\ell-\go t)}.
\end{equation}
Employing the expression for the ellipticity, Eq.~(\ref{ClassEllip}), in
Eq.~(\ref{ClassEx}), we obtain
\begin{equation}\label{ExOutput}
    E^+_x(\ell) \approx
        \mc{E}_x(0)e^{i(k\ell-\go t)}
        \sbr{
            e^{i\gf(0)}
            + g \ell \sin\gf(0)
        }.
\end{equation}

In order to describe squeezing of vacuum fluctuations, one must use a quantum-mechanical
description of the light field.  We introduce the quantum-mechanical operator $\hat{E}_x$
for a monochromatic $x$-polarized light field, which can be written in terms of the
photon annihilation and creation operators, $\hat{a}_x$ and $\hat{a}_x^\dag$,
respectively, as (see, e.g., Ref.~\cite{Loudon87})
\begin{equation}\label{QMefield}
    \hat{E}_{x}
        = \frac{\mc{E}_0}{2} \sbr{\hat{a}_{x} e^{i\prn{kz-\go t}}
        + \hat{a}_{x}^\dag e^{-i\prn{kz-\go t}}},
\end{equation}
where $\mc{E}_0$ is the characteristic amplitude of unsqueezed vacuum fluctuations
\cite{Walls83,Loudon87,Leuchs88,Scully97}. The creation and annihilation operators
satisfy the commutation relations:
\begin{equation}\label{CommutationRelations}
\begin{split}
    \sbr{\hat{a}_\gl,\hat{a}^\dag_{\gl'}} &{}= \gd(\gl,\gl')\\
    \sbr{\hat{a}_\gl,\hat{a}_{\gl'}} &{}= 0,
\end{split}
\end{equation}
where $\gl$ refers to the mode (light frequency and polarization) and $\gd(\gl,\gl')$ is
the Kronecker-delta symbol. When one is interested in the behavior of a multimode field,
one can carry out a sum or integral over the appropriate modes (see, e.g.,
Refs.~\cite{Caves85,Scully97}). Such an approach is important in situations where the
squeezing is frequency dependent over the bandwidth of the light. Here, we assume that SR
does not vary over the bandwidth of the light and that the measurement time is much
longer than the inverse of the spectral width of the SR features, so for simplicity, we
consider a single-mode field and make use of Eq.~(\ref{QMefield}).

Since the degree of squeezing is phase dependent, in an experiment it is necessary to use
a phase-sensitive detection technique (such as a balanced-homodyne scheme
\cite{Loudon87}). In such a scheme one measures the electromagnetic field at a particular
phase $\gc$ with respect to a local oscillator. The phase-dependent electric-field
operator (see, e.g., Ref.\ \cite{Loudon87}) is given by
\begin{equation}\label{QMfield}
    \hat{E}_x(\gc,z)
        = \frac{\mc{E}_0}{2}
        \sbr{
            \hat{a}_x(z) e^{i\gc}
            +\hat{a}_x^\dag(z) e^{-i\gc}
        },
\end{equation}
where we have assumed that the frequency of the local oscillator is the same as that of
the monochromatic light wave, and $\gc$ is the time-independent phase difference between
the local oscillator and the electromagnetic field at the output.

In analogy with the classical formulae for ellipticity, Eqs.~(\ref{ExactEpsilon}) and
(\ref{ClassEllip}), we introduce an ellipticity operator $\hat{\ge}$ for a nearly
$y$-polarized light beam:
\begin{equation}
    \hat{\ge}(z)
        = \mc{E}_0
        \frac{\hat{a}_x(z)-\hat{a}_x^\dag(z)}{2i\mc{E}_y(z)},
\end{equation}
where $\hat{a}_x(z)$ and $\hat{a}_x^\dag(z)$ are the photon annihilation and creation
operators at position $z$. This operator gives the ratio of the component of $E_x$
out-of-phase with $E_y$ to $E_y$. Indeed, supposing that the $x$-polarized field
component is a coherent state $\ket{\gh}$ (which is an eigenstate of the annihilation
operator with eigenvalue $\gh = \abs{\gh}e^{i\gf}$) at position $z$, we can use the above
operator to reproduce Eq.~(\ref{ClassEllip}):
\begin{equation}\label{QM6}
\begin{split}
   \ge
   &{}= \abr{\hat{\ge}}\\
   &{}= \bra{\gh}\frac{\mc{E}_0\sbr{\hat{a}_x(z) - \hat{a}_x^\dag(z)}}{2i\mc{E}_y(z)}\ket{\gh}\\
   &{}= \frac{\mc{E}_0}{2i\mc{E}_y(z)}\prn{\gh-\gh^*} \\
   &{}= \frac{\abs{\gh}\mc{E}_0}{\mc{E}_y(z)}\sin\gf(z)\\
   &{}= \frac{\mc{E}_x(z)}{\mc{E}_y(z)}\sin\gf(z).
\end{split}
\end{equation}

In analogy to Eq.~(\ref{ClassEx}), we have for the $x$-polarized optical electric field
after propagation through the vapor of length $\ell$
\begin{equation}\label{QM7}
    \hat{E}^+_x(\ell)
        = \frac{\mc{E}_0}{2}
        \cbr{
            \hat{a}_x(0)
            + \frac{ig\ell}{2}
            \sbr{
                \hat{a}_x^\dag(0)-\hat{a}_x(0)
            }
        }
        e^{i\gc}.
\end{equation}
It is useful to write the output field operator $\hat{E}_x(\ell)$ in terms of its own set
of creation and annihilation operators as in Eq.~(\ref{QMfield}), in which case the
annihilation operator $\hat{a}_x(\ell)$ at the output of the medium is given by
\begin{equation}\label{OutputAnnihilationOperator}
    \hat{a}_x(\ell)
        = \hat{a}_x(0)
        +\frac{ig\ell}{2}\sbr{\hat{a}_x^\dag(0)-\hat{a}_x(0)}.
\end{equation}

\begin{figure}\label{SqueezedVac_g5}
\includegraphics[width=3.2 in]{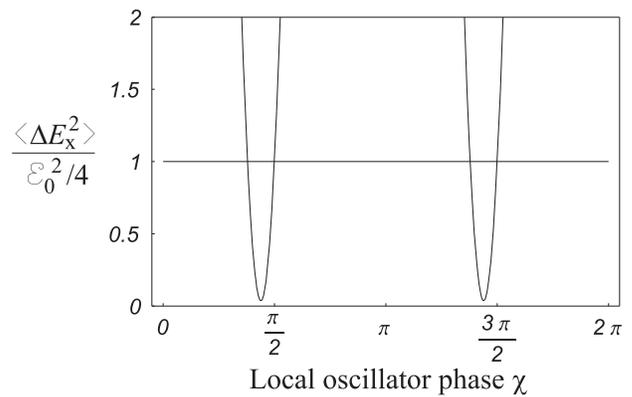}
\caption{Quantum fluctuations $\langle\gD E_x^2\rangle/ \prn{\mc{E}_0^2/4}$ as a function
of the local oscillator phase $\gc$ for the case of zero absorption ($\ga=0$) and $g\ell
= 5$.}
\end{figure}

Let us calculate what happens to the $x$-polarized vacuum field when light that is
linearly polarized along $y$ propagates through the atomic vapor, assuming that there is
no absorption ($\ga=0$). By using our expression for $\hat{a}_x(\ell)$ from
Eq.~(\ref{OutputAnnihilationOperator}) in Eq.~(\ref{QMfield}), we find that the
phase-dependent operator describing the $x$-polarized field after propagation through the
atomic medium is
\begin{widetext}
\begin{align}
\hat{E}_x(\gc,\ell) & = \frac{\mc{E}_0}{2}
        \hat{a}_x(0)\prn{e^{i\gc}-ig\ell \cos\gc} +\frac{\mc{E}_0}{2}
        \hat{a}_x^\dag(0)\prn{e^{-i\gc}+ig\ell \cos\gc}~.
\label{NoAbs3}
\end{align}
\end{widetext}
If the input $x$-polarized field is the vacuum state $\ket{0}$, then the expectation
value of the output field $\langle\hat{E}_x(\gc,\ell)\rangle = 0$. However,
$\langle\hat{E}_x(\gc,\ell)^2\rangle$ is nonzero, so the quantum fluctuations in
$E_x(\gc,\ell)$, given by
\begin{equation}\label{NoAbs5}
\begin{split}
    \abr{\gD E_x(\gc,\ell)^2}
        &{}= \abr{\hat{E}_x(\gc,\ell)^2} - \abr{\hat{E}_x(\gc,\ell)}^2\\
        &{}= \abr{\hat{E}_x(\gc,\ell)^2}\\
        &{}= \frac{\mc{E}_0^2}{4} \prn{1- 2g\ell \sin\gc \cos\gc + g^2\ell^2\cos^2\gc}
\end{split}
\end{equation}
are phase dependent, and for a particular choice of the phase $\gc$ can be made smaller
than the fluctuations of the vacuum field ($\mc{E}_0^2/4$). Fig.~\ref{SqueezedVac_g5}
shows the quantum fluctuations $\abr{\gD E_x^2}/\prn{\mc{E}_0^2/4}$ [Eq.~(\ref{NoAbs5})]
as a function of $\gc$ for the case of $g\ell = 5$, illustrating that there is a
restricted range of phases for which large squeezing is obtained.

The optimum phase of the local oscillator for squeezing is found by setting
\begin{equation}\label{NoAbs6}
  \pdbyd{}{\gc}\abr{\sbr{\gD E_x(\gc,\ell)}^2} = 0.
\end{equation}
We have:
\begin{equation}\label{NoAbs7}
2\cos2\gc + g\ell\sin2\gc = 0,
\end{equation}
so that the optimum phase $\gc_{\text{opt}}$ is
\begin{align}\label{NoAbs8}
    \gc_{\text{opt}} & = \frac{1}{2}\arctan\prn{-\frac{2}{g\ell}} + \prn{n+\frac{1}{2}}\pi
\end{align}
where
\begin{align}
n = 1,2,3,~.~.~.
\end{align}
The minimum fluctuations in the $x$-polarized output field are
\begin{equation}\label{NoAbs9}
   \abr{\sbr{\gD E_x(\gc_{\text{opt}},\ell)}^2}
    = \frac{\mc{E}_0^2}{4}
        \sbr{
            1
            +\frac{g^2\ell^2}{2}
            -\frac{2+g^2\ell^2/2}{\sqrt{1+4/\prn{g^2\ell^2}}}
        }.
\end{equation}
In the limit where $g\ell \gg 1$, we obtain
\begin{equation}\label{NoAbs10}
    \abr{\sbr{\gD E_x\prn{\gc_{\text{opt}},\ell}}^2}
        \approx \frac{\mc{E}_0^2}{g^2\ell^2}.
\end{equation}

In general, since the creation and annihilation operators do not commute
[Eq.~(\ref{CommutationRelations})], there is an uncertainty relationship between the
fluctuations measured at a local oscillator phase of $\chi$ and a phase $\chi+\pi/2$
(see, e.g., Ref.~\cite{Loudon87}):
\begin{equation}\label{QuadratureUncertaintyRelationship}
    \abr{\sbr{\gD E_x\prn{\gc,\ell}}^2}
    \abr{\sbr{\gD E_x\prn{\gc+\pi/2,\ell}}^2}
        \geq \prn{\frac{\mc{E}_0^2}{4}}^{\!2}
\end{equation}
If Eq.~(\ref{QuadratureUncertaintyRelationship}) is satisfied as an equality for some
phase $\chi$, then the electromagnetic field is said to be in a minimum-uncertainty
state.  In this sense, the squeezed vacuum state produced by SR is a minimum-uncertainty
state, as can be seen by choosing $\chi=\chi_text{opt}$ in
Eq.~(\ref{QuadratureUncertaintyRelationship}).

We now turn to SR-media that also have nonzero absorption. One can model a medium with
small total absorption in the following way. We assume that we have an ideal transparent
SR-medium. After passing through this medium, a beamsplitter reflects a small fraction of
the light away. At the same time, vacuum fluctuations enter through the dark port of the
beamsplitter. Thus the squeezed vacuum is attenuated by an amount $e^{-\ga\ell} \approx 1
- \ga\ell$ while $1-e^{-\ga\ell} \approx \ga\ell$ of noise is added in quadrature.
Equation~(\ref{NoAbs10}) becomes
\begin{equation}\label{SmallAbs1}
    \abr{\sbr{\gD E_x\prn{\gc_{\text{opt}},\ell}}^2}
        \approx \frac{\mc{E}_0^2}{4} \prn{\frac{4}{g^2\ell^2} + \ga\ell}.
\end{equation}
We would like to optimize squeezing with respect to $\ell$, so we set
\begin{equation}\label{SmallAbs2}
   \pdbyd{}{\ell}\abr{\sbr{\gD E_x\prn{\gc_{\text{opt}},\ell}}^2} = 0,
\end{equation}
from which we obtain the optimum number of absorption lengths:
\begin{equation}\label{Lopt}
    \ga \ell_{\text{opt}}
        = 2 \prn{\frac{\ga}{g}}^{\!2/3}.
\end{equation}

Substituting the value of $\ell_{\text{opt}}$ from Eq.~(\ref{Lopt}) into
Eq.~(\ref{SmallAbs1}) and taking the limit where $g/\ga \gg 1$, we obtain
\begin{equation}\label{SmallAbs4}
    \abr{\sbr{\gD E_x\prn{\gc_{\text{opt}},\ell_{\text{opt}}}}^2}
        \approx \frac{3\mc{E}_0^2}{4} \prn{\frac{\ga}{g}}^{\!2/3}.
\end{equation}
The squeezing parameter $s$ is the ratio of the amplitudes of the quantum fluctuations
after and before light propagates through the SR-medium:
\begin{equation}\label{SqueezingParm}
   s = \sqrt{\frac{\mc{E}_0^2/4}{\abr{\sbr{\gD E_x\prn{\gc_{\text{opt}},\ell_{\text{opt}}}}^2}}}.
\end{equation}
The reason that $s$ is the relevant parameter describing squeezing is that in order to
take advantage of the reduced noise in the vacuum field, one must interfere the squeezed
vacuum with another field of nonzero amplitude. Then the dominant noise contribution due
to the vacuum appears as a cross term between the real and vacuum field amplitudes. In
such a scenario, $s$ represents the improvement in the signal-to-noise ratio over the
shot-noise limit.

From Eq.~(\ref{SmallAbs4}), we find:
\begin{equation}\label{SqueezeParm}
    s = \frac{1}{\sqrt{3}}\prn{\frac{g}{\ga}}^{1/3}.
\end{equation}
Equation~(\ref{SqueezeParm}) represents a general result for the optimal (with respect to
local oscillator phase $\gc$ and path length $\ell$) vacuum squeezing obtainable in an
SR-medium with small total absorption.

In this article, numerical values for $s$ are given in dB, obtained by taking
$10\log_{10}s$.

Often, it is useful to characterize the light field in terms of the Stokes parameters,
given in terms of the positive and negative frequency components of the field by (see,
e.g., Ref.~\cite{Huard97}):
\begin{equation}
\begin{split}
    S_0 &{}= E_x^+E_x^- + E_y^+E_y^-  \approx \abs{\mc{E}_y}^2\\
    S_1 &{}= E_x^+E_x^- - E_y^+E_y^-  \approx -\abs{\mc{E}_y}^2\\
    S_2 &{}= E_x^+E_y^- + E_x^-E_y^+  \approx \mc{E}_y \prn{E_x^+ + E_x^-}\\
    S_3 &{}= i\prn{E_x^+E_y^- - E_x^-E_y^+}  \approx i\mc{E}_y \prn{E_x^+ - E_x^-},
\end{split}
\end{equation}
so, in the case considered here, we obtain squeezing in particular combinations of the
$S_2$ and $S_3$ Stokes parameters.

\section{Description of the density-matrix calculation}
\label{DMcalc}

In order to determine the vacuum squeezing produced by a given atomic system, we perform
a calculation of the parameters $g$ and $\ga$ based on a standard density-matrix
approach. The time evolution of the atomic density matrix $\gr$ under the action of the
light-atom interaction Hamiltonian $H_L=-\mb{E}\cdot\mb{d}$, where $\mb{E}$ is the
electric field vector, and $\mb{d}$ is the dipole operator, is given by the Liouville
equation (see, e.g., Ref. \cite{Stenholm}):
\begin{equation}
  \frac{d\gr}{dt}=\frac{1}{i\hbar}\sbr{H,\gr}-\frac{1}{2}\cbr{\gG,\gr}+\gL,
  \label{Louiville}
\end{equation}
where the square brackets denote the commutator and the curly brackets the
anticommutator, and the total Hamiltonian $H$ is the sum of $H_L$ and the unperturbed
Hamiltonian $H_0$. $\gG$ is the relaxation matrix (diagonal in the collision-free
approximation)
\begin{equation}
  \bra{\gx Jm}\gG\ket{\gx Jm}=\grg+\grg_0\gd\prn{\gx,\gx_e},
\end{equation}
where $\grg$ and $\grg_0$ are the ground state depolarization rate and the spontaneous
decay rate from the upper state, respectively, and $\gx$ represents the quantum number
distinguishing the ground state ($\gx_g$) from the excited state ($\gx_e$).
$\gL=\gL^0+\gL^{repop}$ is the pumping term, where the diagonal matrix
\begin{equation}
   \bra{\gx_gJ_gm}\gL^0\ket{\gx_gJ_gm}=\frac{\grg\gr_0}{\prn{2J_g+1}}
\end{equation}
describes incoherent ground state pumping ($\gr_0$ is the atomic density), and
\begin{widetext}
\begin{equation}
    \bra{\gx_gJ_gm}\gL^{repop}\ket{\gx_gJ_gm'}
    =\grg_0\sum_{m_e,m_e',q}
        \cg{J_g,m,1,q}{J_e,m_e}
        \cg{J_g,m',1,q}{J_e,m_e'}
        \gr_{\gx_eJ_em_e\gx_eJ_em_e'},
\end{equation}
\end{widetext}
describes repopulation due to spontaneous relaxation from the upper level (see, e.g.,
Ref. \cite{Rautian}). Here $\cg{\ldots}{\ldots}$ are the Clebsch-Gordan coefficients.

The electric field vector is written (see, e.g., Ref. \cite{Huard97})
\begin{equation}\label{lightfield}
\begin{split}
    \mb{E}
        &=\frac{1}{2}
            \sbr{
                E_0e^{i\gf}\prn{\cos\gv\cos\ge-i\sin\gv\sin\ge}e^{i\prn{\go t-kz}}
                +c.c.
            }\hat{x}\\
        &+\frac{1}{2}
            \sbr{
                E_0e^{i\gf}\prn{\sin\gv\cos\ge+i\cos\gv\sin\ge}e^{i\prn{\go t-kz}}
                +c.c.
            }\hat{y},
\end{split}
\end{equation}
where $E_0$ is the electric field amplitude, $\gv$ is the polarization angle, and $\gf$
is the overall phase. By substituting (\ref{lightfield}) into the wave equation
\begin{equation}
  \prn{\frac{\go^2}{c^2}+\frac{d^2}{dz^2}}\mb{E}=-\frac{4\gp}{c^2}\frac{d^2}{dt^2}\mb{P},
\end{equation}
where $\mb{P}=\tr\gr\mb{d}$ is the polarization of the medium, the absorption, rotation,
phase shift, and change of ellipticity per unit distance for an optically thin medium can
be found in terms of the density matrix elements (these expressions are given in Ref.
\cite{budker01}).

\section{Analysis for $J_g=1/2 \ra J_e=1/2,3/2$ systems} \label{SimpleSystems}

\begin{figure}
\includegraphics[width=3.2 in]{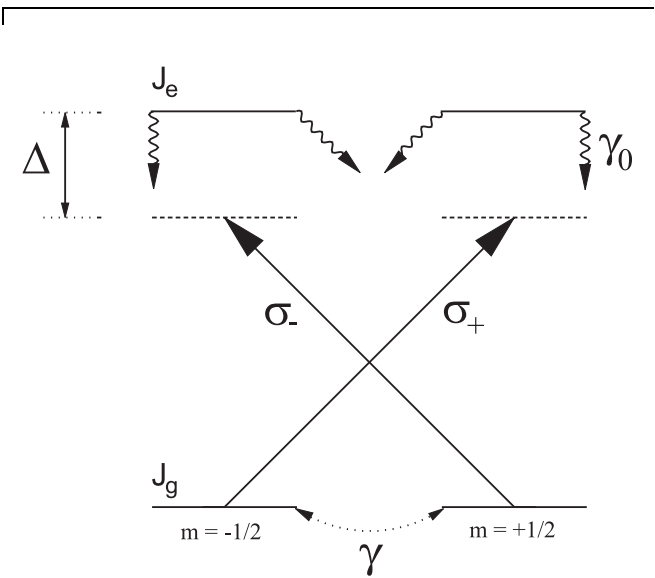}
\caption{Schematic energy level diagram for the X-system ($J_g=1/2 \ra J_e=1/2$
transition). $\grg$ denotes the relaxation rate of ground state atomic polarization;
$\grg_0$ denotes the homogeneous width of the upper state; $\gs_+$ and $\gs_-$ represent
the left- and right-circularly polarized components of the light field, whose frequency
is detuned by $\gD$ from resonance.} \label{Xsystem}
\end{figure}

Using the rotating-wave approximation, we solve Eq.~(\ref{Louiville}) for a closed
$J_g=1/2 \ra J_e=1/2$ transition (an X-system, Fig.~\ref{Xsystem}); analytic solutions
for the optical response are readily obtained in this case.  We assume homogeneous
broadening, since, as will be shown later, the homogeneous width due to power broadening
exceeds the Doppler width under the conditions where large vacuum squeezing is obtained.
This simple system can be realized experimentally in a number of different ways. For
example, at light powers high enough so that hyperfine effects can be neglected, the $D1$
line in alkali atoms can be treated as a simple $1/2\ra1/2$ system. Also, SR in this
system was studied theoretically and experimentally in Ref.~\cite{boyd92}, where K atoms
were employed and He buffer gas was used to collisionally broaden the $4^2S_{1/2} \ra
4^2P_{1/2}$ transition to remove hyperfine effects.

The solutions can be simplified by assuming that $\grg \ll \grg_0$. For the absorption
coefficient $\ga$ as a function of light frequency detuning $\gD$ and in terms of the
optical-pumping saturation parameter
\begin{equation}
    \gk = \frac{d^2 E_0^2}{\grg\grg_0}, \label{kappa}
\end{equation}
where $d$ is the reduced electric-dipole matrix element, we find:
\begin{equation}\label{alpha}
    \ga\prn{\gk,\gD}
        \approx \frac{\ga_0}{1 + 4\prn{\gD/\grg_0}^2 + \prn{\grg/\grg_0}\prn{\gk/3}},
\end{equation}
where $\ga_0=\ga(0,0)$ is the unsaturated absorption coefficient on resonance, given by
\begin{equation}
    \ga_0 = \frac{n}{2\gp} \gl^2 \frac{2 J_e+1}{2 J_g+1},
\end{equation}
where $n$ is the atomic density and $\gl$ is the wavelength of the light. As a function
of $\gD$, $\ga\prn{\gk,\gD}$ is a power-broadened Lorentzian profile. For the
self-rotation parameter $g$, we find:
\begin{equation}\label{SRparm}
    g\prn{\gk,\gD}
        \approx \frac{2}{9} \frac{\ga(\gk,\gD)\gk\gD/\grg_0}{1 + 4\prn{\gD/\grg_0}^2+\gk/9}.
\end{equation}
From Eq.~(\ref{SRparm}), we see that at low light powers ($\gk \ll 1$), SR increases
linearly with light intensity. At high light intensities ($\gk \gg 1$), $g$ falls off as
$\gk^{-1}$.  We also note that while $\ga$ is an even function of detuning, $g$ is
antisymmetric with respect to $\gD$.

From these expressions we find the squeezing $s$ [Eq.~(\ref{SqueezeParm})], optimized in
terms of the sample length $\ell$ and the local oscillator phase $\gc$,
\begin{equation}\label{goa1}
    s\prn{\gk,\gD}
        \approx \frac{1}{3}
                \sbr{
                    \frac{2}{\sqrt{3}}
                    \frac{\gk\gD/\grg_0}{1 + 4\prn{\gD/\grg_0}^2+\gk/9}
                }^{1/3}.
\end{equation}
The detuning that maximizes $s$ for a given $\gk$ is
\begin{align}
\gD_{\text{opt}}=\frac{\grg_0}{2}\sqrt{1+\gk/9}~. \label{Eq:OptDetuning}
\end{align}
The optimum detuning increases with light intensity due to power broadening. Substituting
$\gD_{\text{opt}}$ into Eq. (\ref{goa1}) and taking the limit of large light power gives
the best possible squeezing as a function of light power:
\begin{equation}\label{sOptX}
\begin{split}
    s\prn{\gk,\gD_{\text{opt}}}
        &{}\approx \prn{\frac{\gk}{972}}^{1/6}\\
        &{}= \prn{\frac{d^2E_0^2}{972\grg_0\grg}}^{1/6}.
\end{split}
\end{equation}

The number of unsaturated absorption lengths on resonance required to produce this
squeezing can be obtained using Eq.~(\ref{Lopt}):
\begin{equation} \label{AlphaOpt}
\begin{split}
    \ga_0\ell_{\text{opt}}
    &{}\approx 2\sbr{\frac{\ga(\gk,\gD_{\text{opt}})}{g(\gk,\gD_{\rm opt})}}^{2/3}
        \frac{\ga_0}{\ga(\gk,\gD_{\text{opt}})}\\
    &{}\approx \prn{\frac{4\sqrt{2}}{9}\gk}^{2/3}.
\end{split}
\end{equation}

Performing a similar analysis for a closed $J_g=1/2 \ra J_e=3/2$ transition,
corresponding to the $D2$ line in alkali atoms, one obtains
\begin{equation}\label{sOptW}
\begin{split}
    s\prn{\gk,\gD_{\text{opt}}}
        &{}\approx \prn{\frac{\gk}{7776}}^{1/6}\\
    \ga_0\ell_{\text{opt}}
        &{}\approx \prn{\frac{4\sqrt{2}}{9}\gk}^{2/3},
\end{split}
\end{equation}
which is, up to a numerical factor for $s$, the same as for the X-system.

An important assumption in our derivation of Eq.~(\ref{SqueezeParm}) was that the
ellipticity of the light does not change significantly as the light propagates through
the atomic medium.  Now we can justify this assumption for the cases considered in the
present section. There are two mechanisms which can cause ``self-elliptization'' (SE) of
the light field: linear birefringence and circular dichroism.  In general, because
dichroic effects are related to the absorptive properties of the medium and we are always
in a regime where $\ga\ell$ is small, SE caused by circular dichroism is negligible. In
Ref.~\cite{budker01}, there are two physical mechanisms identified as possible causes of
SR for systems with $J_g=1/2$: optical pumping and ac Stark shifts, both of which lead
only to circular dichroism. Thus SE can be neglected in the present cases. For the
$X$-system, the change in ellipticity $\delta \epsilon$ is described by
\begin{align}
\label{SEparm}
    \frac{\delta \epsilon}{\epsilon(0)\ell}
        \approx \frac{2}{9} \frac{\ga(\gk,\gD)\gk}{1 + 4\prn{\gD/\grg_0}^2+\gk/9}~.
\end{align}
We see that SE is suppressed by a factor of $\gamma_0/\Delta$ compared to SR
[Eq.~(\ref{SRparm})]. Under optimum conditions for squeezing (see
Sec~\ref{SqueezeNumbers}) and $\kappa \gg 1$, $\gamma_0/\Delta \sim 6/\sqrt{\kappa}$ --
confirming that SE is negligible in our case. These arguments also show why in the
present cases SE is not an effective mechanism for producing squeezed vacuum in one of
the circular components of the light field.

\section{Estimate of achievable vacuum squeezing}
\label{SqueezeNumbers}

\begin{figure}
\includegraphics[width=3.375 in]{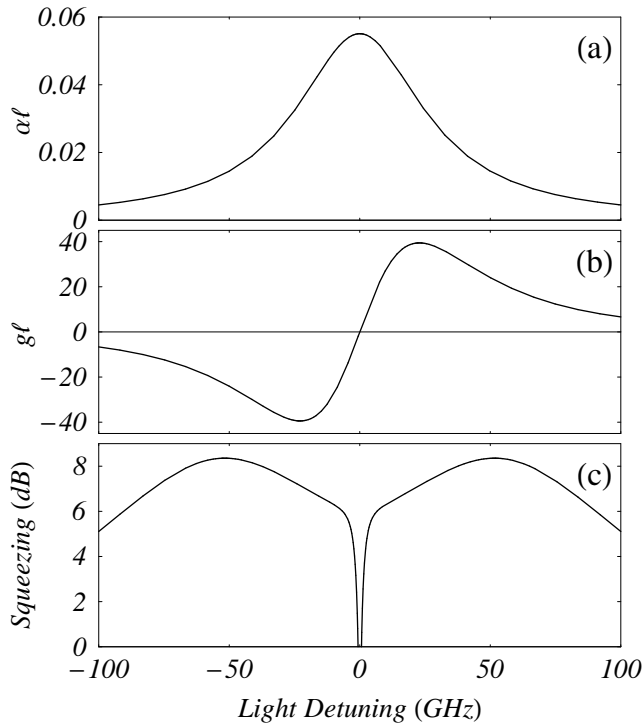}
\caption{Plots of (a) absorption coefficient $\ga$ times path length $\ell$, (b)
self-rotation parameter $g$ [Eq.~(\ref{g})] times path length $\ell$, and (c) vacuum
squeezing in dB for an X-system (Rb $D1$ line at high light power in a buffer-gas-free,
uncoated vapor cell). Light power is 1 W, beam diameter is 100 $\gm$m, atomic density is
$10^{13}~{\rm atoms/cm^3}$, and cell length $\ell = 10$ cm.} \label{XSqueezing}
\end{figure}

First we consider the vacuum squeezing attainable by using a buffer-gas-free cell without
antirelaxation wall coating. Generally, for a given laser power, a smaller beam diameter
results in higher $\gk$ and better squeezing \cite{EndNoteOnBeamDiameter}. For a
reasonably attainable 1-W, 100-$\gm$m-diameter laser beam, the light electric field
amplitude is $\sim$3 kV/cm and the effective ground state relaxation rate resulting from
the transit of atoms through the laser beam is $\sim$3 MHz. For the Rb $D1$ line (which
at these light intensities is effectively an X-system), $\gk \approx 10^8$ and $s \approx
8.4$ dB [Eq.~(\ref{sOptX}), Fig.~\ref{XSqueezing}]. From Eq.~(\ref{AlphaOpt}), we see
that to obtain this level of squeezing, one must use a relatively dense atomic vapor
($\ga_0\ell \approx 2 \times 10^5$, for a cell of length 10 cm, this corresponds to a
density of $\sim$$10^{13}~{\rm atoms/cm^3}$).

Various techniques can be considered to reduce the ground state relaxation rate in order
to produce larger $\gk$, and therefore, in principle, improve vacuum squeezing. Cold atom
traps (in particular far-off-resonant, blue-detuned optical dipole traps) can produce
alkali vapors with ground state relaxation rates $\sim 0.1$ Hz \cite{davidson95}, but the
densities achieved to date are too low to obtain the required optical thickness for
optimum squeezing. In order to achieve $s \simeq 8$ dB with cold alkali atoms, for a
100~$\mu$m long trap (with cross-sectional area of 30~$\mu$m$\times$30~$\mu$m), one would
need $\sim$$10^{10}$ atoms in the trap, which is beyond present capabilities.

Alternatively, buffer gas can be used to reduce the ground state relaxation rate by
lengthening the transit time of atoms through the light beam. The transit time can be
increased until the point where the diffusion rate equals the relaxation rate due to
depolarizing collisions with the buffer gas.  Under conditions where the transit time is
determined by diffusion, the relaxation rate of ground state polarization is
\begin{equation}\label{BufferGasGamma}
    \grg' \approx \frac{\mc{D}}{x^2} + a_1 n_b,
\end{equation}
where $x$ is the light beam diameter, $n_b$ is the buffer gas density, $a_1$ is a
constant describing the rate of depolarizing collisions, and $\mc{D} \approx v/(3 n_b
\gs)$ is the diffusion coefficient ($v$ is the average atomic velocity and $\gs$ is the
cross section for elastic collisions). In addition, one must also take into account the
increase in the homogeneous width of the atomic transition due to pressure broadening.
The homogeneous width is given by
\begin{equation}\label{BufferGasGamma0}
    \grg_0' = \grg_0 + a_2 n_b,
\end{equation}
where $a_2$ is a constant describing the pressure broadening rate. Thus, for a given
light power and beam diameter, the ratio of $\gk'=d^2E_0^2/(\grg_0'\grg')$ to $\gk$ for a
buffer-gas-free cell [Eq.~(\ref{kappa})] is
\begin{equation}\label{KappaPrime}
    \frac{\gk'}{\gk}
    = \frac{\grg_0\grg}{\grg_0'\grg'}
    \approx \frac{\grg_0\grg}{\sbr{v/\prn{3 n_b \gs x^2} + a_1 n_b}\prn{\grg_0 + a_2 n_b}},
\end{equation}
where we note again that $\gk'$ is calculated in the diffusion-limited regime. We see
from Eq.~(\ref{KappaPrime}) that for $n_b$ too large, $\gk'/\gk$ is less than one, so
there is in fact an optimum buffer gas pressure at which squeezing is maximized. For
typical values of cross sections for the relaxation of atomic polarization in
alkali--noble-gas collisions \cite{Franz76} and pressure broadening \cite{Demtroder}, we
estimate the maximum achievable $\gk'/\gk$ to be about 2--3. Thus buffer-gas-filled cells
do not appear to allow substantial improvement in squeezing at high light powers.

Antirelaxation-coated vapor cells can also be used to drastically decrease the relaxation
rate of ground state atomic polarization in the alkalis
\cite{Ensberg,Bouchiat,Alexandrov,UltraNarrow,SensPRA,fmNMOR}. In order to obtain optimal
squeezing in, e.g., a 10~cm diameter paraffin-coated cell, estimates show that the
required atomic density is $n \gtrsim 10^{13}~{\rm atoms/cm^3}$ [Eq.~(\ref{AlphaOpt})].
In this high-density regime, $\grg$ is primarily due to spin-exchange collisions between
the alkali atoms instead of collisions with the wall, which greatly diminishes the
advantages of using paraffin coating. The relatively fast relaxation due to spin-exchange
collisions significantly reduces the probability that an atom retains its polarization
upon returning to the beam after travelling about the cell -- thereby reducing SR related
to coherence effects \cite{budker01}. These factors indicate that one cannot generate
significantly better squeezing via SR by using paraffin-coated cells.

Thus, at present, no system considered above offers significant improvement in vacuum
squeezing via SR compared to buffer-gas-free, uncoated alkali vapor cells.  However, we
note that improvements in the experimental techniques discussed may change this
conclusion in the future.

\section{Calculation for the $^{87}{\rm Rb}$ D-lines} \label{RbDlines}

\begin{figure}
\includegraphics[width=3.375 in]{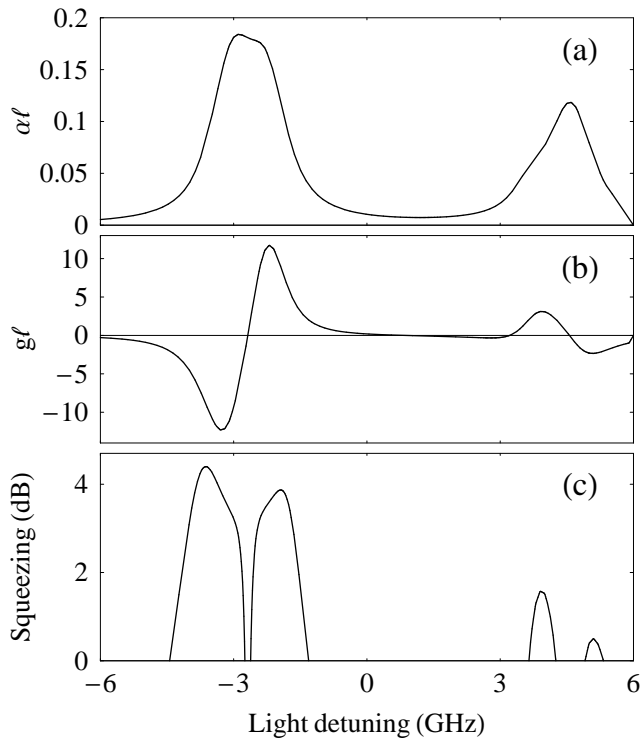}
\caption{Plots of (a) absorption coefficient $\ga$ times path length $\ell$, (b)
self-rotation parameter $g$ [Eq.~(\ref{g})] times path length $\ell$, and (c) vacuum
squeezing in dB for the $^{87}$Rb $D1$ line. Atomic density is chosen to be
$n=10^{12}~{\rm atoms/cm^3}$ to obtain the maximum possible squeezing for the given light
power (10~mW) and light beam diameter (0.03~cm), as described in text.  The Doppler width
is $2\gp \times 306~{\rm MHz}$.} \label{Rb87D1Squeezing}
\end{figure}

\begin{figure}
\includegraphics[width=3.375 in]{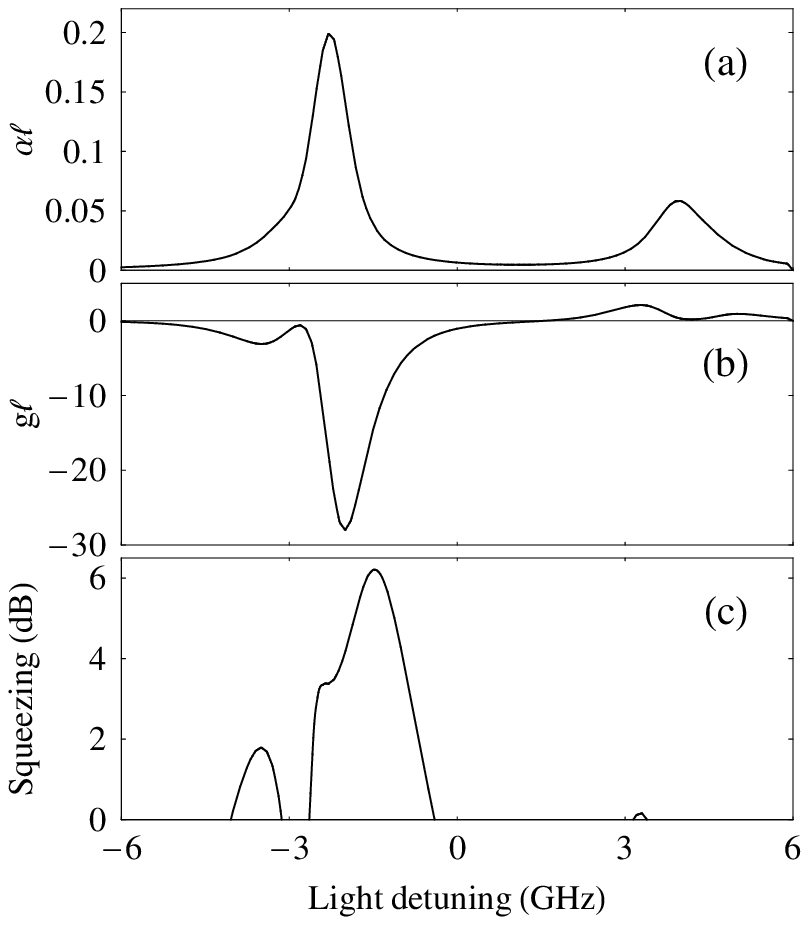}
\caption{Plots of (a) absorption coefficient times path length ($\ga\ell$), (b)
self-rotation parameter times path length ($g\ell$), and (c) vacuum squeezing in dB for
the $^{87}$Rb $D2$ line. Atomic density is chosen to be $n=2\times10^{11}~{\rm
atoms/cm^3}$ to maximize squeezing for the given light power (10~mW) and light beam
diameter (0.03~cm), as described in text. The Doppler width is $2\gp \times 306$
MHz.}\label{Rb87D2Squeezing}
\end{figure}

\begin{figure}
\includegraphics{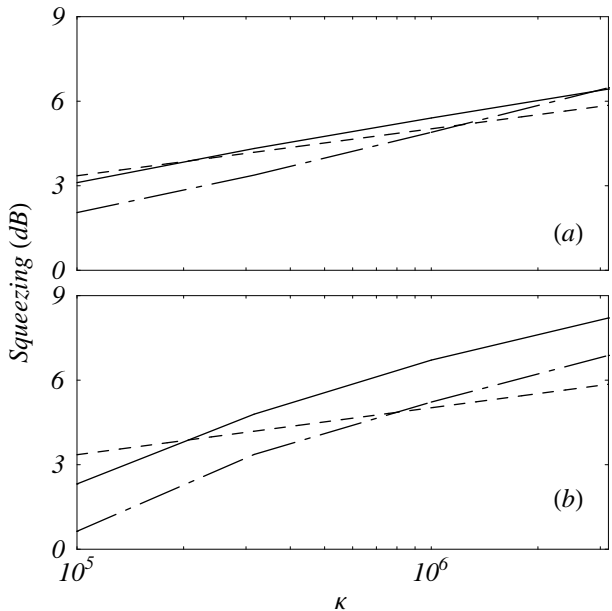}
\caption{Comparison of squeezing as a function of the saturation parameter $\kappa$ for
the X-system (dashed line) to squeezing for the (a) $^{87}$Rb D1 line and (b) $^{87}$Rb
D2 line.  The dot-dashed lines represent squeezing for $F=1 \rightarrow F'$ transitions
and the solid lines represent squeezing for $F=2 \rightarrow F'$ transitions.  The range
of $\kappa$ is chosen so that the ground-state hyperfine levels can be treated
independently, as discussed in the text.} \label{XandRbSqueezing}
\end{figure}

Recent experiments \cite{budker01,novikova00ol} have studied SR in Rb vapors at lower
light intensities than considered in Sec.~\ref{SqueezeNumbers}.  We have performed
density matrix calculations (Sec.~\ref{DMcalc} and Ref.~\cite{budker01}) for conditions
achievable in these experiments. Since our formula for the squeezing parameter $s$
[Eq.~(\ref{SqueezeParm})] is based on the assumptions that $g\ell \gg 1$ and $\ga\ell \ll
1$, we are restricted to light intensities and frequencies where these conditions are
satisfied. In addition, our density matrix calculation treats ground state hyperfine
levels individually. Thus the calculations are not valid for light intensities so high
that the ground state hyperfine structure is not resolved.  In order to use the highest
possible laser intensity in the calculation, we study the $D1$ and $D2$ lines of
$^{87}$Rb, which has larger ground state hyperfine separation than $^{85}$Rb. The results
of these calculations are shown in Figs.~\ref{Rb87D1Squeezing}, \ref{Rb87D2Squeezing},
and \ref{XandRbSqueezing}.

In figures~\ref{Rb87D1Squeezing} and \ref{Rb87D2Squeezing}, we choose laser power equal
to 10~mW (readily obtainable with the tunable diode laser systems employed in
Refs.~\cite{budker01,novikova00ol}) and beam diameter 0.03~cm, which results in the
highest light intensity possible while still resolving the ground state hyperfine lines
($\sim$$10^4~{\rm mW/cm^2}$ \cite{CalcWarning}). Assuming a 10-cm-long vapor cell, we
find the atomic density for which the squeezing is globally maximized (with respect to
light detuning) by finding the light frequency where $g/\alpha$ is maximized, then fixing
the density according to Eq.~(\ref{Lopt}). As can be seen, even with these modest
parameters, squeezing of up to 6~dB can be obtained.

The density-matrix calculations also show that self-ellipticity effects are small for the
$D1$ and $D2$ lines of $^{87}$Rb under these conditions. This verifies that the
assumptions in the derivation of Eq.~(\ref{SqueezeParm}) are satisfied.

At sufficiently high light intensities, the hyperfine effects can be neglected, and the
$D1$ line becomes an X-system, while the $D2$ line becomes a $J_g =1/2 \ra J_e = 3/2$
system, so the analysis of Sec.~\ref{SimpleSystems} applies.  The intermediate regime,
where the hyperfine levels are not fully resolved but hyperfine effects can not be
neglected, is beyond the scope of the present work.  Figure~\ref{XandRbSqueezing} shows
the maximum squeezing as a function of $\kappa$ for the X-system and the $^{87}$Rb D1 and
D2 lines for a range of $\kappa$ where the Rb calculations are valid.  It is interesting
to note that, for this range of $\kappa$, the squeezing in Rb can be somewhat larger than
that obtained for the X-system.

The significant vacuum squeezing that can be obtained in the Rb D-lines at these
relatively low light powers indicates the presence of an enhancement mechanism for
squeezing.  One possible cause of this enhancement is the effect of ground state atomic
coherence, which is not present in the $J_g =1/2 \ra J_e = 1/2$ and $J_g =1/2 \ra J_e =
3/2$ systems considered in Sec.~\ref{SimpleSystems}. Recent research on
electromagnetically induced transparency (EIT) and coherent population trapping has shown
that it is possible to have large nonlinear couplings with very small absorption (see,
e.g., Refs~\cite{WilsonGordon92,lukin98prl,Wong01}) in related systems. Thus, according
to Eq.~(\ref{SqueezeParm}), one could expect that systems where ground state atomic
coherence played a significant role would be favorable for squeezing. There are also SR
mechanisms involving multiple hyperfine transitions \cite{budker01} that play a role in
the Rb D-lines at low light power which are not present for the systems considered in
Sec.~\ref{SimpleSystems}. Further analysis is required to determine the exact physical
mechanism for enhancement of squeezing in this particular case. Recently, in
Ref.~\cite{Mat2002}, vacuum squeezing via SR has been analyzed in the case of the
double-$\gL$ system, a model system in which atomic coherence and interference effects as
well as multi-transition effects play important roles.

Stimulated Raman scattering (SRS) may limit the light intensity which can be used, since
for sufficiently high light intensities, SRS can convert a significant portion of the
incident light power into new frequencies. In addition to the depletion of the input
light beam, nonlinear instabilities associated with SRS may introduce extra noise.
Estimates indicate that the parameters employed in the calculations yielding
Figs.~\ref{Rb87D1Squeezing} and \ref{Rb87D2Squeezing} are in the regime where SRS becomes
significant, so additional analysis is required to determine the manner in which SRS
affects the obtainable squeezing.

In a complete analysis of the noise properties of the light field, in addition to the
absorptive properties of the atomic medium, one must account for the effect of the
quantum noise of atomic states (see, e.g., Refs.~\cite{Polzik99,Kuzmich2000}).  Estimates
indicate that in the high light power, high atomic density regime where the best vacuum
squeezing is obtained (see also Sec.~\ref{SqueezeNumbers}), noise related to these
effects is small enough that it should not degrade the predicted squeezing.

\section{Conclusion}

We have shown that when linearly polarized light propagates through a medium in which
there is nonlinear self-rotation of elliptically polarized light, the vacuum
electromagnetic field in the orthogonal polarization is squeezed.  We have derived a
simple expression [Eq.~(\ref{SqueezeParm})] for the ratio of the amplitude of vacuum
fluctuations at the output of the medium to those at the input under optimum conditions.
Density matrix calculations performed for relatively simple atomic systems indicate that
under realistic conditions it should be possible to achieve $s \approx 8~{\rm dB}$.  We
have compared various experimental systems, such as trapped and cooled atoms and
buffer-gas-filled cells, finding that with presently achievable experimental parameters,
none of the considered techniques offer substantial improvement over buffer-gas-free
vapor cells.  We have also performed calculations showing that vacuum squeezing of up to
6~dB should be achievable in vapors of $^{87}$Rb with the experimental setups used in
Refs.~\cite{budker01,novikova00ol}.

Squeezed vacuum allows one to surpass the shot-noise limit in
polarimetry~\cite{Grangier}.  The proposed technique is a simple way to generate squeezed
vacuum at light wavelengths useful for a variety of applications, such as
magnetometry~\cite{UltraNarrow,SensPRA,fmNMOR,SqueezingInMagnetometry,Phaseonium} and
discrete symmetry tests in atomic systems~\cite{Hunter,Yashchuk,Kimball}.

\acknowledgments

The authors  gratefully acknowledge the support from the Office of Naval Research, the
National Science Foundation, and the Welch Foundation. A.M. acknowledges helpful
discussions with M. O. Scully, M. Fleischhauer and A. S. Zibrov;  D.B., S.M.R. and D.F.K.
would like to thank D. Gauthier, V. Yashchuk, and M. Zolotorev.


\begin{thebibliography}{99}
\bibitem{EndNoteOnBeamDiameter}
From Eqs. (\ref{sOptX}) and (\ref{sOptW}), one can see that for $J_g=1/2 \ra J_e=1/2$ and
$J_g=1/2 \ra J_e=3/2$ transitions, the higher the value of $\gk$ [Eq. (\ref{kappa})], the
better the squeezing.  It turns out that for all the cases considered in this section,
decreasing the light beam diameter either leaves $\gk$ unchanged or makes it larger. This
can be seen as follows. For a given light power, $E_0^2$ scales as $x^{-2}$, where $x$ is
the light beam diameter. The homogeneous width $\grg_0$ is independent of $x$. In all of
the cases we consider here, the relaxation rate of ground state atomic polarization
$\grg$ does not decrease as a function of $x$ faster than $x^{-2}$: $\grg$ is independent
of $x$ for trapped and cooled atoms; in buffer-gas-free cells, the transit rate $\grg_t$
of atoms through the light beam (which can be the dominant contribution to $\grg$) scales
as $x^{-1}$; for cells filled with buffer gas in the diffusion-limited regime, as
discussed below, $\grg_t$ scales as $x^{-2}$. Thus for a given light power, $\gk$ is
either independent of $x$ or is improved by decreasing $x$.
\bibitem{CalcWarning}
This light intensity is 100 times larger than the highest intensity at which our
calculation has been checked against experiment, but we know of no reason why the
calculation should fail under these conditions.
\end{thebibliography}
\end{document}